\documentclass{article}

\usepackage{arxiv}
\usepackage[utf8]{inputenc} 
\usepackage[T1]{fontenc}    
\usepackage{hyperref}       
\usepackage{url}            
\usepackage{booktabs}       
\usepackage{amsfonts}       
\usepackage{nicefrac}       
\usepackage{microtype}      
\usepackage{lipsum}
\usepackage{graphicx}
\graphicspath{ {./images/} }
\usepackage{amsmath}
\usepackage{bm}
\usepackage{mathptmx}
\usepackage{mathrsfs}
\usepackage{orcidlink}

\title{Enhanced Depth Estimation and 3D Geometry Reconstruction using Bayesian Helmholtz Stereopsis with Belief Propagation}

\usepackage{setspace}
\author{
Razieh Azizi \\
 Department of Electrical Engineering,\\
  Amirkabir University of Technology (AUT),\\
  Tehran, Iran \\
  \texttt{r.azizi@aut.ac.ir} \\
   \texttt{razieh.azizi93@gmail.com} \\
   \And
  Hamidreza Amindavar {\textsuperscript{\orcidlink{0000-0002-0954-0674}} } \\
  Department of Electrical Engineering,\\
  Amirkabir University of Technology (AUT),\\
  Tehran, Iran \\
  \texttt{hamidami@aut.ac.ir} \\
  \And
 Hassan Aghaeinia {\textsuperscript{\orcidlink{0000-0002-4286-4662}} } \\
   Department of Electrical Engineering,\\
  Amirkabir University of Technology (AUT),\\
  Tehran, Iran \\
  \texttt{aghaeini@aut.ac.ir} \\
}

\begin{document}
\maketitle
\begin{abstract}
Helmholtz stereopsis is one the versatile techniques for 3D geometry reconstruction from 2D images of objects with unknown and arbitrary reflectance surfaces.  HS eliminates the need for surface reflectance, a challenging parameter to measure, based on the Helmholtz reciprocity principle. Its Bayesian formulation using maximum a posteriori (MAP) probability approach has significantly improved reconstruction accuracy of HS method. This framework enables the inclusion of smoothness priors which enforces observations and neighborhood information in the formulation. We used Markov Random Fields (MRF) which is a powerful tool to integrate diverse prior contextual information and solved the MAP-MRF using belief propagation algorithm. We propose a new smoothness function utilizing the normal field integration method for refined depth estimation within the Bayesian framework. Utilizing three pairs of images with different viewpoints, our approach demonstrates superior depth label accuracy compared to conventional Bayesian methods. Experimental results indicate that our proposed method yields a better depth map with reduced RMS error, showcasing its efficacy in improving depth estimation within Helmholtz stereopsis.
\end{abstract}


\section{Introduction}
3D geometry reconstruction from a set of 2D images has wide applications, including but not limited to virtual reality, computer-aided design, medical visualization, and movie making  \cite{snavely_modeling_2008,tafti_recent_2015}.\\
Various techniques for reconstructing three-dimensional geometry from images have been suggested, each with its own strengths and weaknesses depending on the specific application. One of the main challenges in reconstructing 3D shapes from 2D images is the complexity of estimating surface reflectance. This complexity limits the application of many existing methods to certain types of surfaces, such as Lambertian surfaces. For example, stereo vision-based reconstruction assumes that the surfaces are Lambertian and relies on the presence of sufficient texture for intensity matching of points between different. Another powerful technique is photometric stereo, where a large number of images need to be captured by varying the illumination direction while maintaining a constant viewing direction. This method implicitly requires the surface reflectance to be estimated during the calibration which is a difficult task. Another versatile technique for 3D geometry reconstruction is Helmholtz stereopsis (HS) which eliminates the need for surface reflectance using the reciprocity property and capturing image pairs by exchanging light source and camera position \cite{zickler_helmholtz_2002}.  This method characterizes surface points by both depth and normal from its 2D images without the need for surface reflectance \cite{zickler_helmholtz_2002,roubtsova_bayesian_2014}.\\ 

To improve the performance of HS method for depth and normal estimation several techniques were suggested. For example, radiometric calibration can enhance the accuracy of HS in reconstructed geometry \cite{janko_radiometric_2004}. The problem of applying HS to reconstruct objects with rough and highly textured surface was addressed by \cite{guillemaut_helmholtz_2004}. For reconstruction of real dynamic scenes, use of chromatic calibration procedures in the HS was proposed by Roubtsova and Guillemaut \cite{roubtsova_bayesian_2014,roubtsova_colour_2017}. It was shown that incorporating phase information, a new transpositional reciprocity was suggested which reduces the minimum number of image pairs to one while maintaining the reconstruction accuracy of different surfaces \cite{ding_polarimetric_2023}. The conventional HS solves the maximum likelihood (ML) optimization problem for depth estimation without including the neighbor points and smoothness priors. This causes a sub-optimal solution where the estimated depth map is prone to noise and lacks surface smoothness.\\

A Significant improvement in the HS method has been achieved by employing Bayesian formulation, where the ML problem was replaced by maximum a posteriori (MAP) probability problem. The inclusion of smoothness priors  enforces neighborhood information in the formulation \cite{roubtsova_bayesian_2014,roubtsova_bayesian_2018}. Involving more neighbor points were shown to enhance the depth and normal estimates of HS method \cite{roubtsova_bayesian_2014,roubtsova_bayesian_2018,guillemaut_maximum_2008}. By formulating the HS within a Bayesian framework, depth estimation becomes a labeling problem, enabling the incorporation of constraints derived from prior knowledge and observations. The posterior probability is determined by applying the Bayesian rule, integrating information from both a prior model and a likelihood model. The optimal solution to this labeling problem maximizes a posterior probability (MAP), which is likelihood of a certain value for the variable given the observations \cite{roubtsova_bayesian_2014, li_markov_1994}. This problem is equivalent to minimizing the posterior total energy, which consists of the sum of two terms: the data (or likelihood) energy term and the prior energy term. Three priors were suggested: depth-based, normal based, and a depth-normal prior unifying depth and normal framework to explicitly enforce their consistency \cite{roubtsova_bayesian_2014}. Depth-based prior removes the noise and irregularities by enforcing their smoothness. On the other hand, normal-based prior enforces smooth variations of the normal vectors. Depth-normal prior assumes that the depth and normal are dependent and enforces their consistency.\\

Markov Random Fields (MRF) models are powerful tool to quantitatively integrate diverse prior contextual information or constraints \cite{roubtsova_bayesian_2014}. In MRF framework, which is a undirected graph, each node represents observed or unobserved random variable and each edge represents the relation between two pair variables of the nodes \cite{opipari_dnbp_2024}. The optimal solution of this labeling problem is then equivalent to MAP-MRF problem \cite{li_markov_1994, li_object_2005} where the solution is NP-hard. The reason is that it requires to evaluate all possible labeling configurations and their energy estimation \cite{li_object_2005}. Algorithms based on belief propagation (BP) and graph cuts have been developed highly accurate approximations for these challenging energy minimization problems \cite{boykov_fast_2001}. Both methods share the advantage of identifying local minima which are minimum over large neighborhoods \cite{boykov_fast_2001, sun_stereo_2003}. However, both methods are computationally intensive and have long processing times for solving stereo problems. The running time of belief propagation was reduced by several orders of magnitude, making it suitable to be implemented for vision problems \cite{felzenszwalb_efcient_nodate, kuck_belief_2020}.\\

In this paper, we used MAP-MRF formulation to solve HS problem. We employed the belief propagation method to address the Bayesian Helmholtz problem, leveraging its capacity to include a broader range of neighborhoods, thus enhancing the accuracy of the MAP-MRF method. We used priors based on depth and normal with more neighbor points with which improved the reconstruction accuracy. Additionally, we introduced a new smoothness term within the Bayesian MAP approach, which relies on the normal field integration method for refined depth estimation. Our methodology involved initially calculating the normal vector points obtained from the visual hull method using conventional Helmholtz stereopsis. Subsequently, for each point, we estimated the depth value using the normal field integration method \cite{frankot_method_1988}. Finally, by computing the square of the difference of the calculated depths, we derived the optimal depth. In our approach we included more neighborhoods and the new smoothness normal-based prior which provides stable and accurate geometry reconstruction.

\section{Related Work}
\label{sec:headings}
In this section, we review the main related 3D geometry reconstruction methods from images, discussing their strengths and weaknesses. 

\subsection{3D Shape-from-Silhouettes}
Shape-from-Silhouettes (SfS) is a technique used to compute the visual hull (VH) of an object by intersecting the visual cones obtained from multiple views of the object \cite{baum_novel_1975, laurentini_visual_1994, lazebnik_projective_2007}. Silhouettes refer to the outlines of objects depicted in images. Different techniques, such as image-based, volumetric, surface-based, and hybrid, have been developed to render the VH using the SfS technique.\\ Image-based algorithms construct the VH where all the rendering steps are performed in the image space \cite{matusik_image-based_2000}. The volumetric approach for VH estimation typically subdivides the space into cubes and subsequently determines which cubes belong to the object based on the silhouettes \cite{szeliski_rapid_1993}. Despite the robustness of the volumetric approach for handling complex topologies, there is a trade-off between reconstruction precision and complexity. In other words, the accuracy of voxel based approaches depends on selecting small voxel cells, which in turn increases the model's complexity cubically \cite{liang_3d_2010}. Surface-based, or differential methods, improve the accuracy and performance compared to the volumetric approaches \cite{cipolla_surface_1992}. However, they are sensitive to numerical instabilities compared to the volumetric approaches which are robust for objects with complex shape and topology \cite{liang_3d_2010}. Hybrid methods are suggested to combine the robustness of volumetric and precision of surface-based approaches \cite{boyer_hybrid_2003}.  These methods disregard ineffective cells in volumetric approaches by discretizing the space using sample points lying on the surface of the visual hull, instead of employing regular grids to discretize the space.\\
One of the limitations of Silhouettes based reconstruction methods is that they cannot reconstruct geometries with concavities, as these are not visible in silhouettes. The other challenge is that they require a large number of silhouettes from different view \cite{di_deep_2016, addari_family_2023}.  

\subsection{Stereo-based reconstruction methods - Binocular Stereo and Multi-View Stereo (MVS) }
Among the important three-dimensional reconstruction methods is stereo-based reconstruction, which has been one of the most common methods from the past to the present. Stereo vision deduces the three-dimensional geometry of a scene using two images from different viewpoints \cite{sun_stereo_2003}. In this method, images of an object are captured from different viewpoints under constant lighting. By obtaining information such as the camera positions relative to the object, the distance, and the rotation of the cameras relative to each other, and exploring pixel correspondence of stereo image pairs, the depth of the object can be estimated from the images \cite{moons_3d_2008}. Multi-View Stereo (MVS) vision methods, replicating human eye stereoscopic structure for perception of 3D objects, utilize feature point \cite{ding_polarimetric_2023} matching across images to obtain depth information of the object from disparity map \cite{lin_research_2020, seitz_comparison_2006}. 
The conventional stereo methods rely on the presence of sufficient texture for intensity matching of points between different views to estimate the disparity, which is inversely proportional to depth \cite{seitz_comparison_2006, okutomi_multiple-baseline_1993}.  Disparity map is created by noting deviation of the same point in different images from different observation angles. Although MVS methods, unlike SfS methods, are not confined to the 3D reconstruction of convex objects, their drawback is that they rely on the assumption that the surfaces are Lambertian, i.e., they reflect light proportional to the incoming irradiance and their appearance does not change with viewpoint \cite{ramamoorthi_relationship_2001, scharstein_taxonomy_2001}.
Stereo matching methods face difficulties in exploring pixel correspondence of stereo image pairs due to noise, texture-less, or occluded regions of images \cite{sun_stereo_2003, liu_adaptively_2024}. The Bayesian approach addresses several of these challenges in stereo matching by encoding prior constraints such as smoothness and uniqueness in the problem and treating the problems as an inference problem.\\
Markov Random Fields (MRFs) provide a versatile tool to encode dependencies between variables and model spatial interactions. Therefore, MRFs can be incorporated into a Bayesian framework to establish prior distributions and direct posterior inference. This formulates Bayesian stereo matching as a maximum a posteriori MRF (MAP-MRF) problem \cite{sun_stereo_2003}. Belief Propagation (BP) which is an efficient algorithm to solve inference problems can be applied to find optimal solution of MAP-MRF problem \cite{sun_stereo_2003,zhao_mrf_2023}. 

\subsection{Photometric stereo}
Photometric stereo is another powerful image based 3D reconstruction technique \cite{woodham_photometric_1980}. In this method images are captured by varying the illumination direction while the viewing direction is unaltered by holding the camera location fixed. Under this condition, image intensities can be obtained, which can subsequently provide the surface orientation of points needed to reconstruct the 3D geometry. Photometric stereo demonstrates good performance for surfaces with smooth and uniform properties. One of the difficulties of Photometric stereo lies in its need to capture images of an object under large number of various illuminations \cite{vogiatzis_practical_2010}. In addition, although photometric stereo does not explicitly rely on the known reflectivity, it is required in the calibration phase, which is a difficult task \cite{roubtsova_bayesian_2014}.

\subsection{Helmholtz Stereopsis}
Combining stereo method and photometric stereo, Helmholtz stereopsis (HS) was introduced which is a versatile technique for 3D geometry reconstruction \cite{zickler_helmholtz_2002}. Helmholtz stereopsis is a technique for reconstructing the 3D geometry of an object by characterizing surface points by both depth and normal from its 2D images without the need for surface reflectance \cite{zickler_helmholtz_2002, roubtsova_bayesian_2014}. Helmholtz stereopsis distinguishes itself from other geometry reconstruction techniques, such as stereo vision, structure from motion, and photometric stereopsis, by its ability to operate independently of the bidirectional reflectance distribution functions (BRDFs) of objects—a measure used to quantitatively describe surface reflectance \cite{nicodemus_geometric_1977}. This independence constitutes its primary strength and advantage. To initialize the HS technique, a rough 3D outline of the object, i.e. its visual hull, is needed. SfS method for visual hull estimation combines well with the HS method since neither requires knowledge of the reflectance function.

\subsubsection{Reciprocal image pairs and Helmholtz reciprocity}
\begin{figure}
  \centering
  \includegraphics[width=7cm]{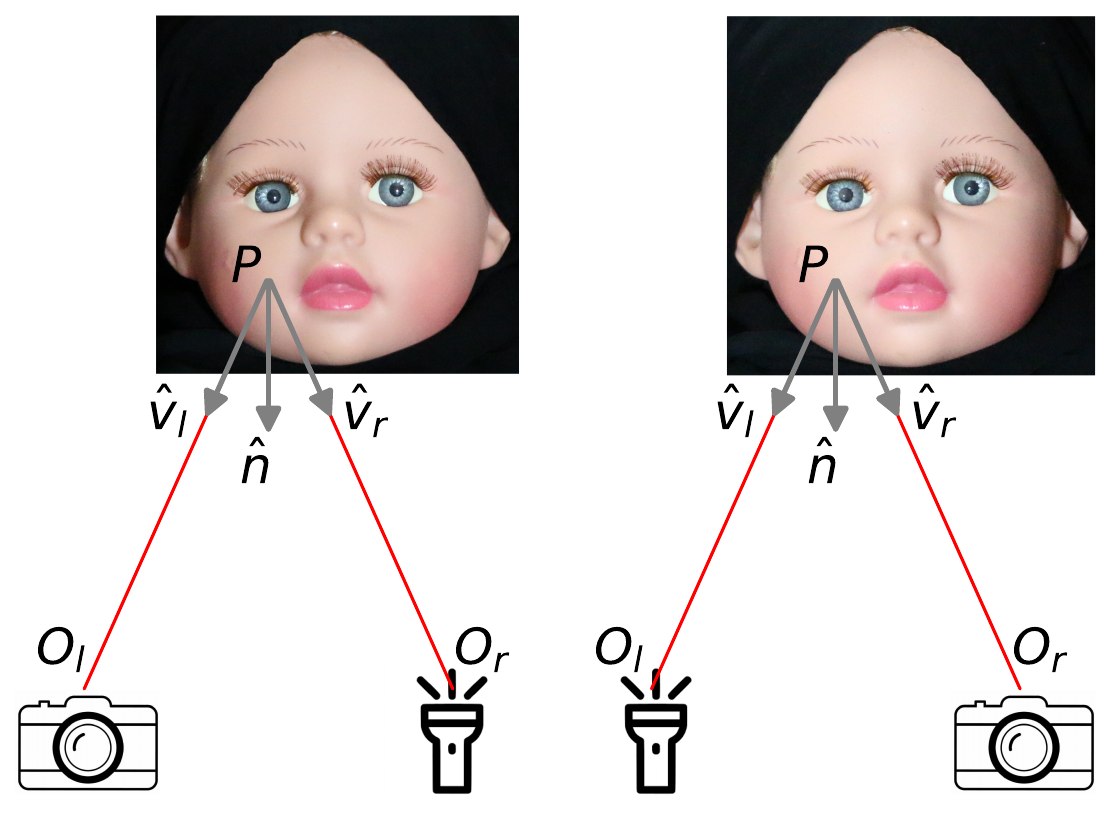}
  \caption{The imaging method of objects using the Helmholtz stereopsis technique: the camera and the light source are placed in the desired location and the first image is captured. Then the position of the camera and the light source are swapped, and the second reciprocal image is taken.}
  \label{fig:fig1}
\end{figure}

The BRDF of a point on a surface is denoted by \( f_r(\hat{i}, \hat{e}) \) and represents the ratio of the outgoing radiance to the incident irradiance, where the first and second arguments correspond to the directions of the incident and reflected light rays, respectively. In order to illustrate the relationship between the radiation values in the corresponding image points of reciprocal image pairs, consider Figure \ref{fig:fig1}. The left image in  Figure \ref{fig:fig1} shows the imaging procedure for an object with a light source positioned at \( O_r \) and a camera at \( O_l \). For a point \( P \) positioned on the surface of the geometry, the direction from \( P \) to the light and camera, denoted by unit vectors \( \hat{\bm{v}}_r \) and \( \hat{\bm{v}}_l \), respectively, are
\begin{equation}
    \hat{\bm{v}}_r = \frac{O_r - P}{|O_r - P|} \, , \quad
    \hat{\bm{v}}_l = \frac{O_l - P}{|O_l - P|}. 
    \label{eq:1}
\end{equation}

Image irradiance, which specifies the brightness or intensity of light on a pixel \cite{ikeuchi_determining_1990}, at the projection of \( P \) is given by

\begin{equation}
    I_l = f_r(\hat{\bm{v}}_r, \hat{\bm{v}}_l) \frac{\hat{n} \cdot \hat{\bm{v}}_l}{|O_r - P|^2} \, , 
    \label{eq:2}
\end{equation}

where \( \hat{\bm{n}} \) is the unit normal vector to the surface at point \( P \). Therefore, the inner product in Eq. (\ref{eq:2}) provides the cosine of the angle between the direction to the light source and the surface normal. Similarly, in the reciprocal case where the locations of the light source and camera are exchanged, the image irradiance is expressed as
\begin{equation}
    I_r = f_r(\hat{\bm{v}}_l, \hat{\bm{v}}_r) \frac{\hat{n} \cdot \hat{\bm{v}}_l}{|O_l - P|^2} \, . \label{eq:3}
\end{equation}

Due to Helmholtz reciprocity, the BRDF is equal for two corresponding points of two images, that is
\begin{equation}
    f_r(\hat{\bm{v}}_r, \hat{\bm{v}}_l) = f_r(\hat{\bm{v}}_l, \hat{\bm{v}}_r). \label{eq:4}
\end{equation}

It is worth emphasizing that in image pairs of conventional stereo methods, which are captured with a fixed light source, this relationship does not hold. This is the advantage of the Helmholtz stereopsis method, which allows combining Eqs. \eqref{eq:2} and \eqref{eq:3} to eliminate the BRDF to obtain

\begin{equation}
    \left(I_l \frac{\hat{\bm{v}}_l}{|O_l - P|^2} - I_r \frac{\hat{\bm{v}}_r}{|O_r - P|^2}\right) \cdot \hat{\bm{n}} = \bm{W}(d) \cdot \hat{\bm{n}}.
    \label{eq:5}
\end{equation}

In the above equation, the normal \( \hat{\bm{n}} \) and the depth \(d\) are unknown.  The values of \(I_l \) and \(I_r \) are images intensities. From geometrically calibrated camera and light source,  \(O_l \) and \(O_r \) are known, and values of \(\hat{\bm{v}}_l \), \(\hat{\bm{v}}_r \)  and \(P\) are obtained from a geometric calibrated camera \cite{hartley_multiple_2003, heikkila_four-step_1997, zhang_flexible_2000}. Eq. \ref{eq:5} is obtained from a single pair of reciprocal images and has three unknowns: two for normal \( \hat{\bm{n}} \) and one for depth \(d\). Therefore, at least three reciprocal pairs are required to solve the equations to find these unknowns. In the next section, binocular stereopsis is generalized to a n-view stereopsis \cite{zickler_helmholtz_2002, barros_opto-mechanical_2007} to establish the framework for Helmholtz method.  

\subsubsection{Helmholtz Stereopsis method}
Generalizing the binocular constraint (Eq. \ref{eq:3}) to the multinocular case provides the required constraints to estimate the depth value \( d \). In this approach, \( n \) cameras are located at \( O_c \) (\( c = 1, \ldots, n \)) together with a principal camera at \( O_p \) which is utilized to parametrize the depth search. The principal camera might be selected at one of the \( O_c \) locations. A point \( q \) is assumed on one of the images which gives a set of \( n \)-points \((q_1, \ldots, q_n)\) for \( n \)-images. 

A set of possible discrete values for \( d \) (\( d \in D = \{d_0, \ldots, d_{N_D}\} \)) is assumed, and the goal is to find the best value representing the correct image depth for a given set of image intensities recorded at \( n \)-point sets \( \bm{Q}(d) = \{q_c(d), c = 1, \ldots, n\} \). With \( N_p \) pairs of captured reciprocal images, it is possible to establish \( N_p \) linear constraints sorted in matrix \( \bm{W}(d) \in \mathbb{R}^{N_p \times 3} \), in which its \( j \)-th row is defined by:

\begin{equation}
    \bm{w}_j(d)^T = \left( I_{l_j} \frac{\hat{\bm{v}}_l}{|O_{l_j} - P|^2} - I_{r_j} \frac{\hat{\bm{v}}_r}{|O_{r_j} - P|^2} \right)^T .
    \label{eq:6}
\end{equation}

Note that \( (O_{l_j}, O_{r_j}), j=1, \ldots, N_p \) are captured from different pairs of positions. Using these \( N_p \) pairs, the constraint in Eq. (5) can be written as:

\begin{equation}
    \bm{W}(d) \hat{\bm{n}} = 0
    \label{eq:7}
\end{equation}

The above equation represents the Helmholtz radiometric constraint, which has been utilized as a matching constraint in the work by \cite{guillemaut_helmholtz_2004,guillemaut_maximum_2008}. 

To determine the depth, \(d\), and the unit normal vector, \(\hat{n}\), from the above equation, singular value decomposition (SVD) can be applied to \(W\) \cite{roubtsova_bayesian_2014}. That is

\begin{equation}
W = UDV^T,
\label{eq:8}
\end{equation}

where \(U\) and \(V\) are unitary matrices, and \(D\) is a diagonal matrix with elements \(\sigma_3 \leq \sigma_2 \leq \sigma_1\). The superscript \(T\) denotes the transpose. The last column of \( V \) provides the normal at the sampled point \cite{roubtsova_bayesian_2014}. In the absence of noise, the rank of \( W(\bar{d}) \) will be 2 for a correct depth value, \(\bar{d}\) \cite{zickler_helmholtz_2002}. However, in many cases, noise is present, and therefore the rank of \( W \) becomes 3. In other words, when the mutual constraint is fulfilled, \(\sigma_3\), the last diagonal value of the matrix \( D \), approaches zero. 

To search for the correct depth value, a complementary measure for correct points located on the surface of the geometry (i.e., depth search) is the ratio of the second to the third diagonal value, \(\sigma_2 / \sigma_3\), expressed as:

\begin{equation}
r_q(d) = \frac{\sigma_2}{\sigma_3}.
\label{eq:9}
\end{equation}

When dealing with accurate depth values, the above ratio tends to be large \cite{zickler_helmholtz_2002, roubtsova_bayesian_2014}. To obtain a stable solution for accurate depth that satisfies Eq. \ref{eq:7}, one strategy is to assume that the depth of the surface remains locally constant around a given point \( q_0 \). Therefore, the search for the correct depth should include points within a small rectangular window \( N \) around \( q_0 \), as:

\begin{equation}
d_q^* = \underset{d \in \mathscr{D}}{\text{arg max}} \sum_{q \in \mathscr{N}} r_q(d).
\label{eq:10}
\end{equation}

Subsequently, the corresponding normal to the obtained depth is estimated using 

\begin{equation}
\hat{\bm{n}}^*_q = \underset{d \in \mathscr{D}}{\text{arg max}} \|\bm{W}_q(d^*)\hat{\bm{n}}\|^2, \|\hat{\bm{n}}\|=1.
\label{eq:11}
\end{equation}

The above HS formulation does not explicitly enforce consistency between depth and normal estimates. Depth estimation in HS has inherent ambiguity. In the conventional HS, the \textit{maximum likelihood} (ML) is used to estimate individual depth values independently of their neighbors, which results in non-smooth surfaces and a lack of fine details in the reconstructed geometry due to measurement noise, calibration uncertainties, discretization errors, and sensor saturation \cite{roubtsova_bayesian_2014, roubtsova_bayesian_2018}. Additionally, while in the HS formulation the surface points are uniquely characterized by both depths and normals, ML depth estimation does not exploit the consistency of depth and normal during the estimation or integration stage.

\subsubsection{Bayesian approach for depth and normal estimation as maximum a posteriori (MAP) optimization in Helmholtz stereopsis }
Bayesian statistics provides a robust framework for addressing reconstruction problems by modeling the measurement process and prior assumptions as probability distributions \cite{jenke_bayesian_2006}. One of its main advantages is the ability to encode various prior constraints, such as spatial smoothness and uniqueness \cite{sun_stereo_2003}.  Using the Bayesian approach, the goal is to achieve best reconstruction of the scene structure S from the measurement D, which is created from the S with including uncertainties such as statistical error. Probability of a reconstruction S based on the given measurement D is obtained as:

\begin{equation}
P(S \mid D) = \frac{P(D \mid S) \, P(S)}{P(D)}.
\label{eq:12}
\end{equation}

Having information from measurement system, a model is assumed for \(P(D\mid S) \), representing probability distribution of the likelihood of measurements D being made of scene structure S. Bayesian statistics approach requires prior distribution, which is probability distribution \(P(S)\) over the set of all possible original scenes. Having exact probabilistic model for \(P(S)\) is not feasible and therefore usually partial prior knowledge, such as assumption of smooth surfaces, is assumed for \(P(S)\). Using a Bayesian framework for estimating each point's depth value in the HS method enables the incorporation of information from neighboring points (Bayesian priors). The use of the Bayesian framework improves accuracy of geometry reconstruction by effectively incorporating statistical dependencies among neighboring depth estimates which enforces continuity of depth map \cite{roubtsova_bayesian_2014}. This can be characterized as a labeling problem, incorporating constraints derived from prior knowledge and observations. This involves assigning a depth label, \(z(x,y)\), to each random variable \(p(x,y) \), allowing its formulation within a Bayesian framework. 

Bayesian reconstruction techniques heavily rely on prior probabilities, as these define what is considered noise and what should be regarded as object features \cite{jenke_bayesian_2006}.  Subsequently, Bayes’ rule is used to find best \(P(S \mid D)\) as given in Eq. \ref{eq:12}. This involves maximizing a posterior probability (MAP), \(P(S \mid D)\), through numerical optimization and the procedure is called maximum a posteriori solution \cite{jenke_bayesian_2006, han_review_2017}.  

Markov Random Fields (MRFs), as a branch of probability theory, serve as a framework for analyzing spatial or contextual dependencies in physical phenomena. In the context of Helmholtz Stereopsis (HS) for geometry reconstruction, MRFs provide a robust framework to model spatial relationships between neighboring points (i.e., pixels) and play a crucial role in computational vision for geometry reconstruction \cite{li_markov_1994, szeliski_comparative_2008}. Using an MRF formulation, each pixel is considered a node, and the surface normal at each point is treated as a random variable. The edges between these nodes represent the dependencies between neighboring pixels. MRFs model the probabilistic relationships between these variables, incorporating constraints such as local smoothness and continuity of the surface.

By formulating the HS problem using Markov Random Fields (MRFs) within a Bayesian framework, the optimal solution can be defined through Maximum a Posteriori (MAP) estimation, which aims to determine the most probable labeling of the surface normals given the observed data.
Achieving an optimal solution to the labeling problem is equivalent to energy minimization, where the energy function consists of two terms: one term incorporates the observed data, and the other term enforces spatial relationships between neighboring points. The solution can be accomplished by maximizing the a posteriori probability (MAP). The depth estimation of Helmholtz stereopsis as a MAP problem can be formulated as \cite{roubtsova_bayesian_2014}:

\begin{align}
f^*_{\text{MAP}} &= \underset{f \in S}{\text{arg max}} \sum_{p(x,y)} \Big[ (1 - \alpha) E_d(x, y, z_{p(x,y)}) \notag \\
&  + \sum_{p(x',y') \in N(p(x,y))} \alpha E_s(x, y, z_{p(x,y)}, x', y', z_{p(x',y')}) \Big].
\label{eq:13}
\end{align}

In the above equation, parameter \(\alpha \) determines the balance between  \(E_d\) and \(E_S \), the data and smoothness terms, respectively. Since the MRF optimization is formulated as a minimization problem, the appropriate distribution for it is the Gibbs distribution, which is formulated as

\begin{equation}
E_d(x, y, z) = e^{-\mu \times \frac{\sigma_2(v(x, y, z))}{\sigma_3(v(x, y, z))}}.
\label{eq:14}
\end{equation}

where \(\mu = 0.2 \times \log(2)\) is the decay factor \cite{roubtsova_bayesian_2014} and the function is constrained within the range \(0\) to \(1\) as it indicates likelihood. Note that for the correct depth value, \(\frac{\sigma_2}{\sigma_3}\) tends to infinity (see Eq. \ref{eq:9}). Both data and smoothness terms in the energy function are defined for each random variable \(p(x, y)\) and given \(p({x'},{y'}) \in N(p(x, y))\). The space \(\mathscr{N}(p(x, y))\) is the Markovian neighborhood of \(p(x, y)\) within the visual hull (VH) defined as \cite{roubtsova_bayesian_2014}.

\begin{equation}
\mathscr{N}(p(x, y)) = \{ p(x + k, y + (1 - k)) \mid k \in \{0, 1\}, \, R_{p(x + k, y + (1 - k))}^V \neq \emptyset \}.
\label{eq:15}
\end{equation}

Incorporating the smoothness prior in the MAP-MRF optimization problem, as outlined in Eqs. (\ref{eq:13})--(\ref{eq:15}) prevents sub-optimal solutions and yields depth estimates that are more accurate and less noisy. Three types of smoothness terms -- Depth-based, Normal-based, and Depth-normal consistency prior -- were defined by Roubtsova and Guillemaut \cite{roubtsova_bayesian_2014} to generate both depth and normal estimates by Helmholtz stereopsis method. Depth-based priors remove the noise and irregularities by enforcing the smoothness between depth values. Normal-based priors enforce the gradual spatial variations of the normal. Notably, among these terms, the Depth-normal consistency prior demonstrated superior performance.

\section{Methodology - Improved depth and normal estimation in HS}
To improve the accuracy of depth and normal estimates in Helmholtz-Stereopsis, we suggest including more neighbor nodes and suggest new smoothness functions. In addition, we use Belief propagation to solve the MAP-MRF optimization problem.

\subsection{Inclusion of more neighbor nodes in smoothness energy functions for the MAP optimization problem}

For the first modification, we adopt the normal-based prior smoothness function proposed in \cite{roubtsova_bayesian_2014}, but modification of incorporating additional neighbor points. Roubtsova and Guillemaut \cite{roubtsova_bayesian_2014} estimated Normal-based prior using only the upper and right nodes, resulting in inferior performance compared to other energy functions. The reason for the weaker performance can be attributed to not using all the information in the neighboring nodes. In an image, the neighboring pixels typically have similar reflectance which results in having normal vectors with similar alignment and direction. This is the constraint used in the photometric stereo method \cite{woodham_photometric_1980}, where the estimated unit normal are proportional to the reflectance of the pixels. Therefore, we incorporate the four neighboring nodes and the adjacent nodes of each neighboring and then estimate the sum of the internal angles between their unit normal vectors. A group of nodes exhibiting a smaller sum of the internal angles, signifying greater similarity, will possess nearly aligned normal vectors. This observation implies that the desired point is accurately positioned at the correct depth. In \cite{roubtsova_bayesian_2014} the smoothness term is

\begin{equation}
E_{s_1} = \pi^{-1} \arccos(\hat{\bm{n}}_p, \hat{\bm{n}}_q).
\label{eq:16}
\end{equation}

where \( p \) is the point for which its depth is being estimated, and \( q \) is the neighboring point. Their unit normal vectors, obtained by solving the Helmholtz stereopsis method, are denoted by \( \hat{\bm{n}}_p \) and \( \hat{\bm{n}}_q \), respectively.

\subsection{Proposed smoothness function based on normal field integration method}

The second smoothness function is a proposed function that uses the normal field integration method \cite{frankot_method_1988}. In this method, we calculate the normal vector points  using the conventional Helmholtz, then we calculate the depth z for all three-dimensional points with normal field integration method \cite{frankot_method_1988}. Then, by using the square of the difference of the calculated  z , we estimate the optimal depth and define the smoothness term as follows.

\begin{equation}
E_{s_2}(z_p, z_q) = (z_{in-p} - z_{in-q})^2.
\label{eq:17}
\end{equation}

MAP Helmholtz stereopsis problem is formulated like Eq. \ref{eq:13} with different smoothness function, as

\begin{align}
Z^*_{MAP} &= \underset{z \in S}{\text{arg max}} \sum_{p(x,y)} \left[ (1 - \alpha) E_d(z_p) \right. \nonumber \\
& + \sum_{q(x,y) \in \mathscr{N}(p(x,y))} \alpha E_s(z_p, z_q) ] \label{eq:18}
\end{align}

In the above equation, \(\mathscr{N}(p(x,y)\) are neighbors that are obtained with belief propagation approach which is explained in the next section.

\subsection{Helmholtz stereopsis problem based on the MAP optimization using belief propagation method}

Expressing the Helmholtz stereopsis as a labeling problem, for each random variable \( p(x, y) \), a depth label \( z_{p(x, y)}^* \) is assigned. The goal is to find the best labeling (true labels) that satisfies the spatial constraints in the model. The optimal solution to this labeling problem maximizes the posterior probability (MAP), which represents the likelihood of a certain value for the variable given the observations \cite{roubtsova_bayesian_2014, li_markov_1994}. This problem is equivalent to posterior energy minimization, where the total energy consists of the sum of two terms: the data (or likelihood) energy term and the prior energy term.

The Markov random field framework well suits modeling such spatial structures and interactions. In MRF, which is a undirected graph, each node represents observed or unobserved random variable and each edge represents the relation between two pair variables of the nodes \cite{opipari_dnbp_2024}. The optimal solution of this labeling problem is then equivalent to MAP-MRF problem \cite{li_markov_1994, li_object_2005}. Finding the global optimal solution to Eq. \ref{eq:18} is NP-hard as it requires to check all the possible labeling configurations and estimation of their energy \cite{li_object_2005}. Extensive research has been conducted to develop sampling and variational based methods to approximate these difficult-to-solve problems \cite{kuck_belief_2020}.  In this case, Belief Propagation (BP), which is a variational method, is an efficient algorithm for finding the optimal solution to inference problems, such as MAP paired with MRF \cite{sun_stereo_2003}.  BP involves an iterative exchange message algorithm within the network between neighboring nodes of the graph. BP computes the joint probability of the graph by using the edges as propagation path and passing message between nodes \cite{li_object_2005, sun_stereo_2003}. BP encompasses two algorithms: Max-product and Sum-product. However, for MAP-MRF problem the BP algorithm is equivalent to the iterative Max-product, a message update rule of belief propagation \cite{li_object_2005, zhao_mrf_2023}. This algorithm is employed to determine the most likely configuration of variables within the graph, commonly referred to as the maximum a posteriori (MAP) estimate. In the context of current work, Max-product message passing algorithm \cite{weiss_optimality_2001} is well-suited for finding the minimum of the energy function. 

Assume \( R_{p(x, y)} = \{ v(x, y, z) \mid x = x^*, y = y^* \} \) represents the set of voxels where we want to find the optimum depth \( z_{p(x, y)} = z_{p(x, y)}^* \) for each random variable \( p(x, y) \). In this problem, the labels are depth values shown by \( Z = \{ z_1, \dots, z_N \} \) assigned to the two-dimensional \((x, y)\) points. In other words, if \( P \) denotes the set of two-dimensional random variables, then for each \( p \in P \), a depth value \( z \in Z \) is assigned.

The MAP problem is approached as a propagation algorithm \cite{sun_stereo_2003}, with the goal of determining the correct depth label for point \( q \). If \( p \) is one of the neighbors of \( q \), the message that \( p \) sends to \( q \) in the iteration \( t \) is:

\begin{equation}
m_{p \to q}^{(t)} = \min_{f_p} \left( E_s(z_p, z_q) + E_d(z_p) + \sum_{S \in \mathscr{N}(p) / q} m_{s \to p}^{(t-1)}(z_p) \right)
\label{eq:19}
\end{equation}

where  \(\mathscr{N}(p)\) denotes the neighbors of \(p\) except \(q\) and the message values are zero for \( t = 0 \), i.e., \( m_{p \to q}^{0} = 0. \). The data term, \( E_d \), was defined in Eq. (\ref{eq:14}). Increasing the number of iterations \( T \) involves more neighbors in the process. For a given point \( q \), it is necessary to calculate all four messages received from its neighbors on the right, left, up, and down. Following this, for each of these messages, a label depth \( z_p \) is selected based on the iteration \( T \). After the completion of iteration \( T \), all these messages are aggregated by the data term of point \( q \), and a belief vector is computed for each node.

\begin{equation}
b_q (z_q) = E_d(z_p) + \sum_{p \in N(q)} m_{p \to q}^T(z_p).
\label{eq:20}
\end{equation}

The equation above is computed for all labels of point \( q \), and ultimately, the label is chosen based on the minimum belief function. Representing all the belief functions of point \( q \) as a set \( B_q = \{ b_q (z_{q1}), \dots, b_q (z_{qn}) \} \) where \( n \) is the number of labels assigned to point \( q \). The correct depth label is obtained as:

\begin{equation}
z_q^* = \underset{z_q \in Z}{\text{arg min}} \, B_q(z_q).
\label{eq:21}
\end{equation}

\begin{figure}[h]
\centering
  \includegraphics[width=7cm]{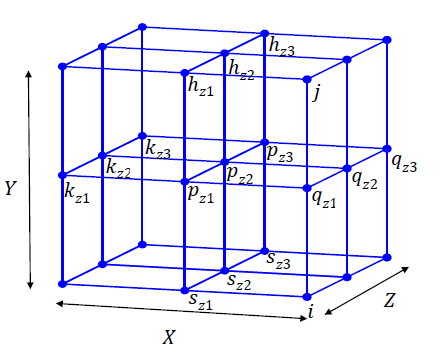} 
\caption{A set of 3D voxels showing neighboring points of a voxel along with their corresponding labels for selecting the appropriate depth using the belief propagation method.}
\label{fig:2}
\end{figure}

An example will serve as an illustration. A small set of three-dimensional voxels is shown in Figure \ref{fig:2}. To solve the Helmholtz stereopsis, a space similar to the one shown in Figure \ref{fig:2} should be considered, which encompasses the actual object. Solving the Helmholtz stereopsis equation for all the voxels yields a unit normal and a value \( r_q \) (Eq. \ref{eq:9}), constituting the data term. For each point in the \( xy \) plane, only one of the available \( z \) values is selected. Following this, the surface is constructed by utilizing the normal vectors of these points and adopting a unifying normal approach.

\par %
Now we want to select the correct depth value for point \( q(x, y) \) among the values of the labels \( p_{z1}, p_{z2}, \ldots, p_{zN} \). The message received by point \( q \) from its neighbors for \( t = 1, 2, \ldots \) is defined as:

\( t = 1 \):
\begin{equation*}
m_{p \to q}^1 (z_q) = \min_{z_p} \left\{ E_s (z_p, z_q) + E_d (z_p) \right\}
\end{equation*}

\( t = 2 \):
\begin{align}
m_{p \to q}^2 (z_q) &= \min_{z_p} \left\{ E_s (z_p, z_q) + E_d (z_p) + \right. \nonumber \\
& \quad \left. \min_{z_h} \left( E_s (z_h, z_p) + E_d (z_h) \right) + \right. \nonumber \\
& \quad \left. \min_{z_k} \left( E_s (z_k, z_p) + E_d (z_k) \right) + \right. \nonumber \\
& \quad \left. \min_{z_s} \left( E_s (z_s, z_q) + E_d (z_s) \right) \right\}
\label{eq:22}
\end{align}

\( t = 3 \):
\begin{equation*}
\vdots
\end{equation*}

As \( T \) increases, more neighbors are added, and after \( T \) iterations, the value \( z_p \) for point \( p \) is selected, and a message is sent to point \( q \). From all neighbor points \( i \) and \( j \) near \( q \) that send messages to point \( q \), the belief function of point \( q \) for all its labels is computed as:

\begin{equation}
b_q (z_q) = E_d (z_q) + m_{p \to q}^T (z_q) + m_{j \to q}^T (z_q) + m_{i \to q}^T (z_q)
\label{eq:23}
\end{equation}

And finally, among the three computed beliefs for \( q \), a label is selected whose belief is minimum:
\begin{equation}
z_q^* = \underset{z_q}{\text{arg min}} \left\{ b_q (z_{q1}), b_q (z_{q2}), b_q (z_{q3}) \right\}
\label{eq:24}
\end{equation}

\section{Results}
In this section, Helmholtz stereopsis method is used for geometry reconstruction. Depth values and their normal unit vectors are estimated by minimizing the energy of MAP-MRF optimization problem. We used three different positions for placing the camera and the light source because three reciprocal image pairs are needed, each acquired by exchanging the positions of the light source and the camera. Initially, these three positions are calibrated both geometrically and radiometrically \cite{janko_radiometric_2004, heikkila_four-step_1997}. Three distinct locations for the camera and light source are utilized to capture the three image pairs. The camera was a Canon EOS 7D, paired with a CFL light source (23W, 230V) emitting white light. No additional light sources were present in the imaging room. To enhance image-background separation, a black background was employed. The distance between the camera and the object was maintained at 130 cm.\\
 To assess the performance of the suggested method for geometry reconstruction, we employed various objects with polished and glossy surfaces, along with rough surfaces. Additionally, our goal was to observe the method's outcomes on surfaces with both high and low complexities.

\subsection{Initial geometry for Helmholtz stereopsis method}
The Helmholtz stereopsis (HS) method requires an initial 3D volume to reconstruct the geometry. A rough 3D outline of the object is estimated using the Shape-from-Silhouette (SfS) method \cite{laurentini_visual_1994, lazebnik_projective_2007}. The SfS method for visual hull estimation complements the HS method since neither requires knowledge of the reflectance function. In the SfS method, the visual hull of the object is derived from the intersection of visual cones obtained from multiple views.\\
To separate the object of interest from the background, we convert the images into binary images so that the object and background are white and black, respectively. This is performed by defining a threshold for the intensity of images. To facilitate this procedure, the background was darkened in the image-capturing process. Subsequently, we consider a volume of three-dimensional points, commonly referred to as voxels encompassing the object. Using the projection matrix of each camera, obtained through calibration, we transform the three-dimensional points of assumed volume into two-dimensional points and then project them onto the images. 
The silhouette method selects the points that are present in all images on the object of interest and removes the remaining points. Then, it converts the selected points into three-dimensional space and creates a visual hull of the object  
For an accurate reconstruction of object using the silhouette method, it is necessary to have a large number of images taken from various angles around the object. However, in this context, we employ the Silhouette method to obtain an initial and rough approximation of the object, which serves as input for the Helmholtz stereopsis method. Using the reconstructed visual hull voxels and camera location parameters, Unit vectors \(\hat{\bm{v}}_r \) and \(\hat{\bm{v}}_l \) are calculated according to Eq. (\ref{eq:1}).

\subsection{Helmholtz stereopsis reconstruction method}
\begin{figure}
  \centering
  \includegraphics[width=8cm]{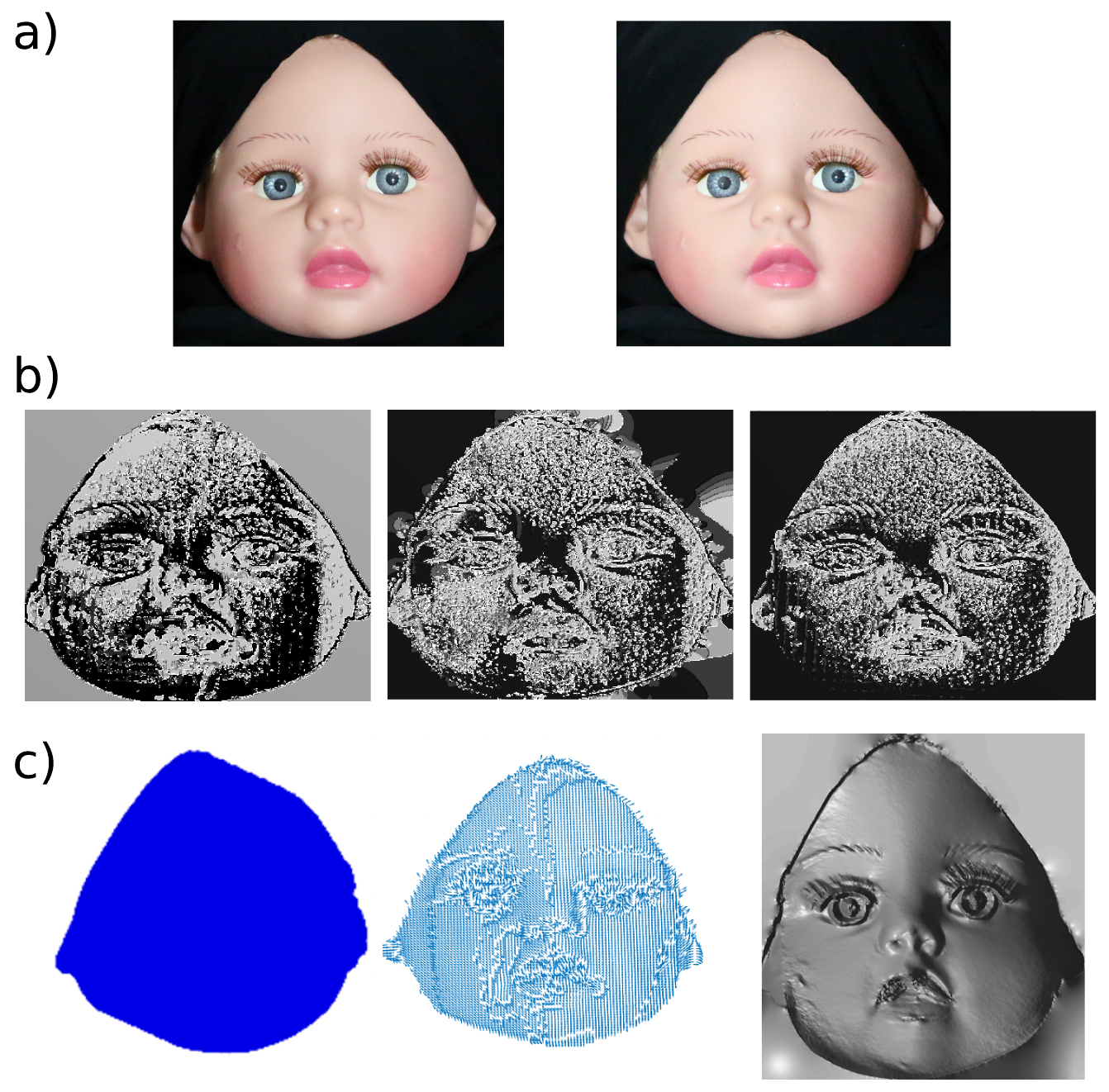}
  \caption{a) One image pair from three pairs of stereo images of a doll’s face, taken by exchanging the camera and light source. b) Depth estimates by three methods, MAP-ND (left), MAP-BP-N (middle), and MAP-BP-Z (right). c) Visual hull is shown in the left subplot, normal vectors and the 3D reconstructed geometry of doll’s face are shown in the middle and right subplots, respectively.}
  \label{fig:fig3}
\end{figure}

\begin{figure}
  \centering
  \includegraphics[width=8cm]{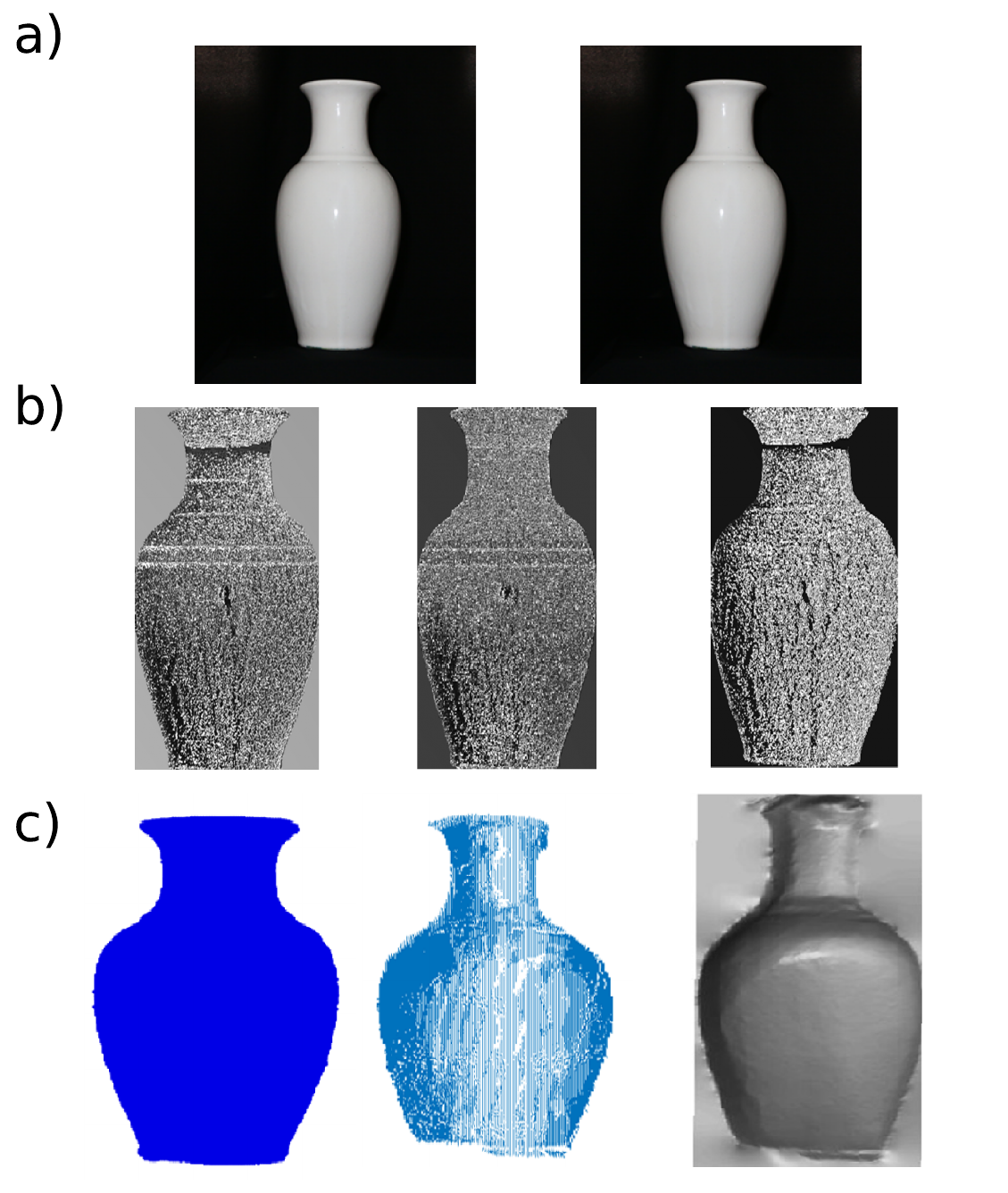}
  \caption{a) One image pair from three pairs of stereo images of a vase, taken by exchanging the camera and light source. b) Depth estimates by three methods, MAP-ND (left), MAP-BP-N (middle), and MAP-BP-Z (right). c) Visual hull is shown in the left subplot, normal vectors and the 3D reconstructed geometry of vase are shown in the middle and right subplots, respectively.}
  \label{fig:fig4}
\end{figure}

\begin{figure}
  \centering
  \includegraphics[width=8cm]{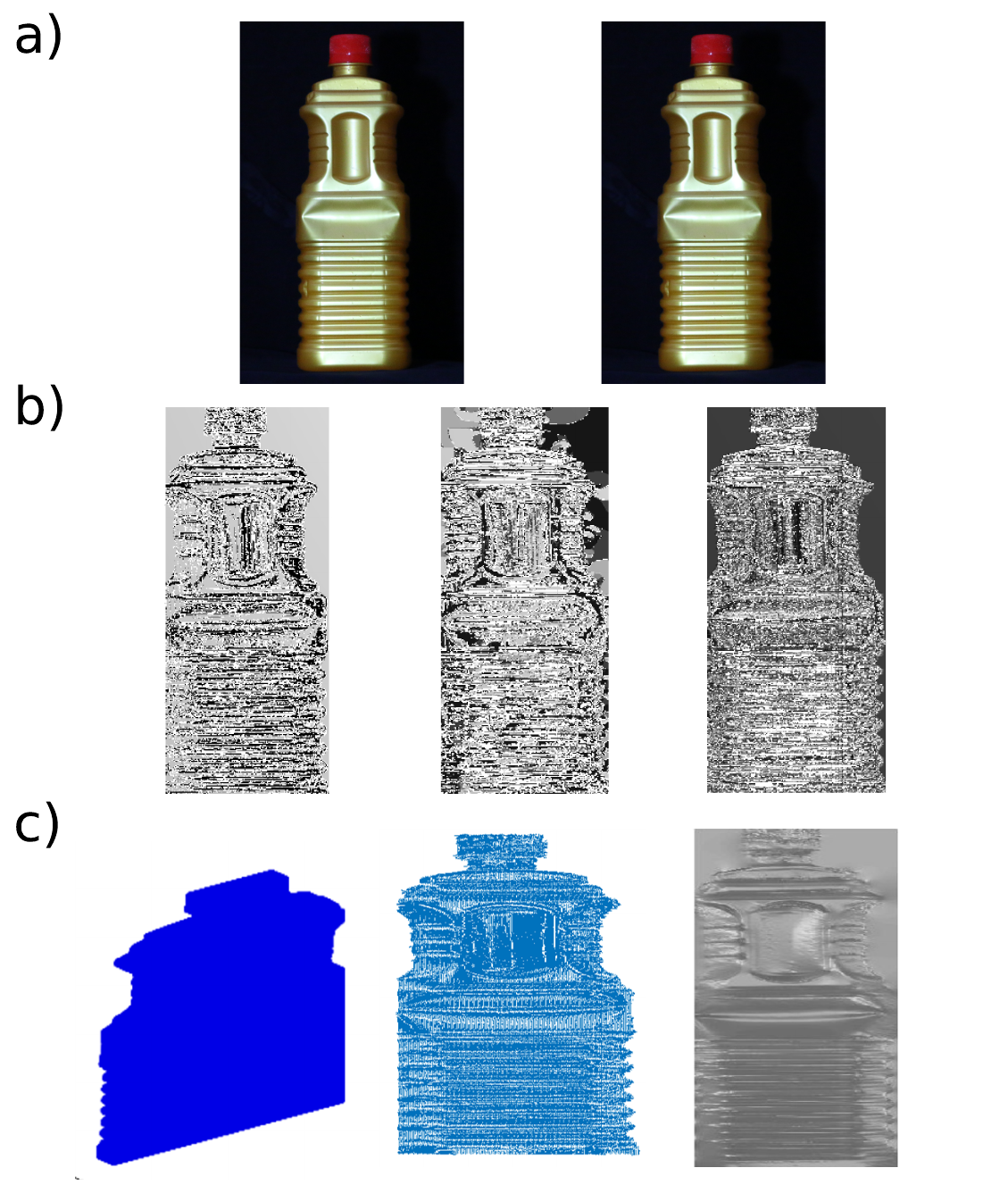}
  \caption{a) One image pair from three pairs of stereo images of an oil bottle, taken by exchanging the camera and light source. b) Depth estimates by three methods, MAP-ND (left), MAP-BP-N (middle), and MAP-BP-Z (right). c) Visual hull is shown in the left subplot, normal vectors and the 3D reconstructed geometry of oil bottle are shown in the middle and right subplots, respectively.}
  \label{fig:fig5}
\end{figure}

\begin{figure}
  \centering
  \includegraphics[width=8cm]{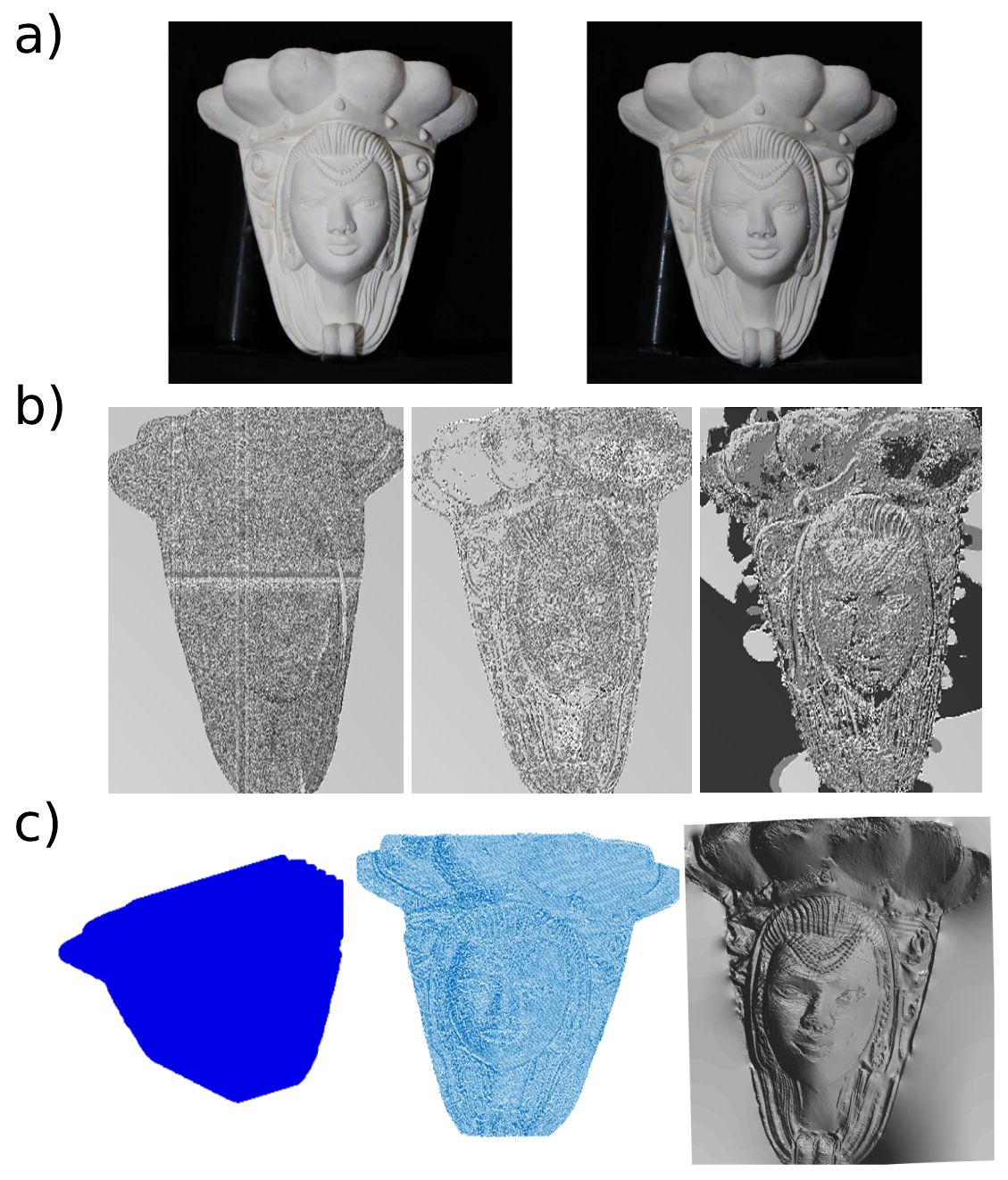}
  \caption{a) One image pair from three pairs of stereo images of a gypsum sculpture, taken by exchanging the camera and light source. b) Depth estimates by three methods, MAP-ND (left), MAP-BP-N (middle), and MAP-BPZ (right). c) Visual hull is shown in the left subplot, normal vectors and the 3D reconstructed geometry of gypsum sculpture are shown in the middle and right subplots, respectively.}
  \label{fig:fig6}
\end{figure}

In the Helmholtz stereopsis method, each point within the three-dimensional space of the visual hull corresponds to a point present in all three image pairs. The intensities of these points in the images, along with the unit vectors directing from the camera to the corresponding points on the visual hull, are calculated for application in the Helmholtz stereopsis equation as per Eq. (\ref{eq:6}). The intensities are calculated after applying a Gaussian filter to the images, which not only smooths the intensities but also reduces variations in intensity among adjacent points caused by noise. To estimate the depth \( d \) and two components of the normal vector \( \hat{\mathbf{n}} \), at least three equations are required. Three reciprocal pairs of images are used to form Eq. (\ref{eq:7}) for each point. By applying SVD (see Eq. (\ref{eq:8})) to this set of equations, we estimate the normal vector of the point and parameter \( r_q \) (see Eq. (\ref{eq:9})) which is used for estimating the depth of the point. To reconstruct the object, we need to estimate the depth of each point together with its corresponding normal vector. Then the surface of the object is reconstructed using the integration normal vectors approach. We estimated the depth using three methods:
\begin{enumerate}
    \item Bayesian method with smoothness function based on depth-normal MAP-N \cite{roubtsova_bayesian_2014},
    \item Bayesian method based on belief propagation with smoothing function based on normal MAP-BP-N,
    \item Bayesian method based on belief propagation with smoothing function based on depth obtained from the normal field integration method MAP-BP-Z.
\end{enumerate}
The performance of suggested methods is evaluated in reconstructing the few different objects with different surfaces and materials. These objects include a doll’s face (Fig.  \ref{fig:fig3}), a vase (Fig.  \ref{fig:fig4}), an oil bottle (Fig. \ref{fig:fig5}), and a gypsum sculpture (Fig.  \ref{fig:fig6}). Three reciprocal image pairs are captured for each object by exchanging the positions of the light source and camera at three different perspectives. 

The depth maps and the final reconstruction results for objects indicate that we have been able to reconstruct details with high precision using the minimum pairs of images in the Helmholtz stereo method, which consists of three pairs of images. For example, in the image of the doll’s face, the Helmholtz method was able to reconstruct eyebrows and even eyelashes correctly. For the gypsum sculpture, with intricate surface details and complexity compared to other objects, which due to its matte surface is less sensitive lighting variations compared to glossy surfaces, the reconstruction was reasonably good.

To evaluate the closeness between the estimated depth and the final reconstructed depth using normal vectors, root mean square errors are calculated, as:

\begin{equation}
e_{\text{RMS}} = \sqrt{\frac{\sum_{i \in P} \left(z_{t_i} - \hat{z}_i\right)^2}{N_P}} \quad 
\label{eq:25}
\end{equation}

where \( z_{t_i} \) is the final reconstructed depth obtained by unifying normal vectors, and \( \hat{z}_i \) is the estimated depth for point \( i \). The set \( P \) denotes the pixels used in the reconstruction, and \( N_P \) represents the total number of pixels.

The \( e_{\text{RMS}} \) values for different objects, computed using the three methods, are normalized within the range of depth values (i.e., \( 0 \leq z \leq 1 \)). These values are presented in Table \ref{tab:1}.

\begin{table}[ht]
\centering
\caption{Normalized \( e_{\text{RMS}} \) values for different objects using three depth estimation methods.}
\label{tab:1}
\begin{tabular}{|c|c|c|c|c|}
\hline
\textbf{Depth Estimation Method} & \textbf{Doll’s Face} & \textbf{Vase} & \textbf{Oil Bottle} & \textbf{Gypsum Sculpture} \\
\hline
MAP - BP - Z & 0.38 & 0.33 & 0.41 & 0.43 \\
MAP - BP - N & 0.41 & 0.45 & 0.45 & 0.62 \\
MAP - ND & 0.46 & 0.47 & 0.48 & 0.83 \\
\hline
\end{tabular}
\end{table}

After comparing the \( e_{\text{RMS}} \) values in Table \ref{tab:1} for all four studied objects, it is evident that the MAP-BP-Z method exhibits the least error compared to the other two methods, indicating superior reconstruction accuracy. For objects with lower complexity and smoother surfaces, such as vases, doll’s faces, and bottles, the performance of the MAP-BP-Z method is slightly better than the other two methods. 

However, for surfaces with higher complexity, such as gypsum sculptures, the difference in relative estimated error is significantly smaller for the MAP-BP-Z method compared to the other two methods. Gypsum sculptures, in particular, exhibit the highest complexity compared to other surfaces. Additionally, due to the material being gypsum, which lacks polish and possesses a complex reflectance function, the depth estimation process is more intricate.

Observing the depth maps for the three methods, it is evident that the MAP-BP-Z method yields better results compared to the other two methods. The MAP-ND method shows an almost nonexistent depth map. In contrast, the MAP-BP-N method provides improved results compared to the MAP-ND method, but it still lacks the detail observed in the MAP-BP-Z method. Overall, the proposed MAP-BP-Z method demonstrates superior performance compared to the simpler Bayesian MAP-ND method. Specifically, among the proposed methods, MAP-BP-Z yields more accurate results than the MAP-BP-N method.

\section{Discussion and conclusions   }

Helmholtz stereopsis is a versatile method for reconstructing the 3D geometry of objects with various materials independent of their surface reflectance. For instance, it outperforms conventional stereo methods in reconstructing texture-less smooth surfaces. Moreover, Helmholtz stereopsis excels in reconstructing surfaces with complex textures or rough surfaces that do not effectively reflect light rays, whereas the photometric method faces challenges in reconstructing the geometry of such surfaces.\\

In this paper, we have enhanced the precision of depth estimation in Helmholtz stereopsis (HS), which is required for determining the normal vector used in the subsequent reconstruction of geometry. The initial depth estimation in the Helmholtz stereopsis method is usually not very accurate. However, having a more precise initial depth estimation leads to a more accurate selection of the normal vector and, consequently, a higher accuracy in the final three-dimensional reconstruction.

We demonstrated that depth estimation using a Bayesian approach based on belief propagation provides a more accurate depth map. This improved accuracy is achieved since the correct depth is selected by including the neighbors, which provides better depth estimates for all the points. A new smoothness term was suggested for the Bayesian problem. The smoothness term, based on the minimum differences of neighbor depths obtained by unifying the normal vectors, demonstrated better performance compared to the one suggested in \cite{roubtsova_bayesian_2014}, which uses the angle between the normal vectors of the neighbors.

This result indicates that using the inner angle between the normal vectors of the point of interest and its neighbors cannot be used for correct depth estimation. The reason is that the angle between the vectors cannot uniquely show the direction of the normal vector. For example, if a vector has an angle \(\alpha \) relative to another vector, it does not determine whether the second vector is on the right or left side of the first vector. Although we cannot overlook the similarity of the normal vectors created for pixels with the same intensity, the angle between them is not sufficient to determine their direction.

Smoothness term based on unifying the normal vectors calculates the gradients of normal vectors and selects an appropriate depth value for the normal vector of interest by including the neighbors. We demonstrated that using the proposed smoothness terms and solving the Bayesian problem by belief propagation to select the correct depth values provides enhanced depth map for complicated, rough, and textured surfaces. Including more neighborhoods and the new smoothness normal-based prior estimated equally or slightly more accurate geometry reconstruction compared to the depth-normal prior.  As a result, the root mean square (RMS) error for this method, between the estimated depth and the final reconstructed depth, is lower compared to the two previous methods. It is worth noting that using the least number of image pairs, the method could reconstruct geometric details with high accuracy. For example, in the case of the doll’s face, features such as eye lines, eyelashes, and eyebrows are accurately reconstructed.


\bibliographystyle{unsrt} 
\bibliography{references} 








\end{document}